\documentclass{PoS}

\usepackage{wrapfig}
\usepackage{amssymb}

\title{An update on nuclear PDFs at the LHeC}

\ShortTitle{An update on nuclear PDFs at the LHeC}

\author{\speaker{Hannu Paukkunen for the LHeC study group} \\
University of Jyvaskyla, Department of Physics, P.O. Box 35, FI-40014 University of Jyvaskyla, Finland \\
Helsinki Institute of Physics, P.O. Box 64, FI-00014 University of Helsinki, Finland \\
Instituto Galego de F\'\i sica de Altas Enerx\'\i as (IGFAE), Universidade de Santiago de Compostela, E-15782 Galicia, Spain \\
E-mail: \email{hannu.paukkunen@jyu.fi}}

\abstract{The prospects for a measurement of nuclear parton distribution functions (PDFs) at the Large Hadron--Electron Collider are discussed in the light of recent progress made in the front of global analysis of nuclear PDFs.}

\FullConference{XXV International Workshop on Deep-Inelastic Scattering and Related Subjects\\
		3-7 April 2017\\
		University of Birmingham, UK}

\begin{document}

\section{Introduction}

\begin{wrapfigure}{r}{0.43\textwidth}
\vspace{-1.2cm}
\center
\includegraphics[width=0.43\textwidth]{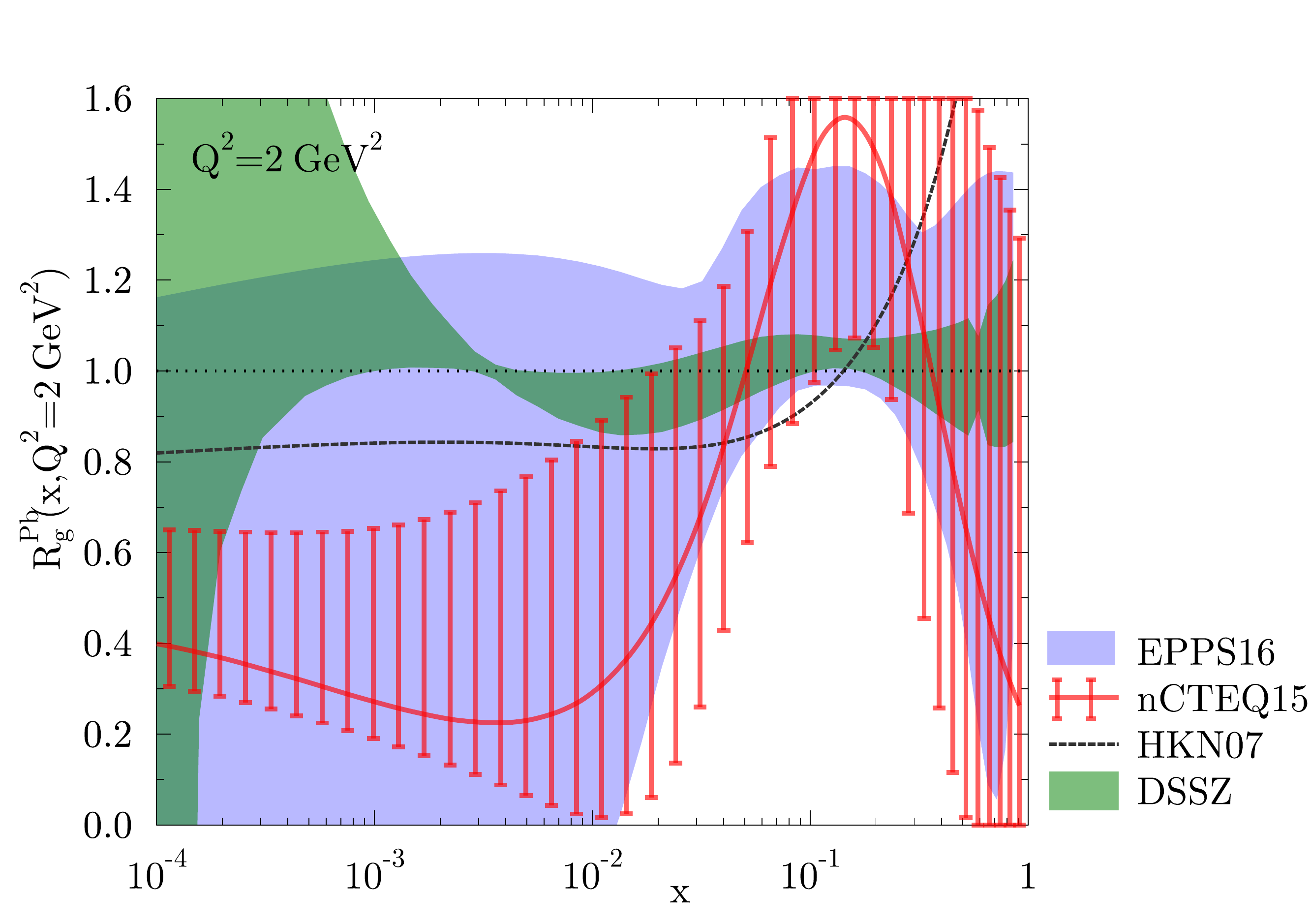}
\vspace{-0.8cm}\caption[]{A comparison between the gluon nuclear modifications at $Q^2=2 \, {\rm GeV}^2$ from the EPPS16 \cite{Eskola:2016oht}, DSSZ \cite{deFlorian:2011fp}, nCTEQ15 \cite{Kovarik:2015cma}, and HNK07 \cite{Hirai:2007sx} global analyses of nuclear PDFs.}
\label{fig:nPDFs}
\end{wrapfigure}

\vspace{-0.3cm}
One of the key missions of the planned Large Hadron--Electron Collider (LHeC) \cite{AbelleiraFernandez:2012cc} is a precision determination of the nuclear PDFs. The prevailing situation in the case of gluon nuclear modification, defined as a ratio $R_g^{Pb} = f_g^{\rm Pb}/f_g^{\rm p}$ between the bound-proton gluon PDF $f_g^{\rm Pb}$ and that of the free proton $f_g^{\rm p}$, is illustrated in Figure~\ref{fig:nPDFs}: especially at low values of momentum fraction $x$ (meaning here, $x \lesssim 10^{-2}$) and smallish scale $Q^2$, the behaviour is essentially unknown. The large uncertainties may become a bottleneck for e.g. progress in theoretical understanding of the origin of nuclear-PDF effects, distinguishing signatures of non-linear evolution, precision studies of phenomena in heavy-ion collisions (which often involve a small, but still perturbative scale), and calculations of cosmic-ray interactions in the air relevant for neutrino telescopes. Such physics reasons underscore the importance of a precision determination of nuclear PDFs.

\vspace{-0.3cm}
\section{Capabilities of LHC in constraining small x}

\begin{wrapfigure}{r}{0.45\textwidth}
\center
\vspace{-1.5cm}
\includegraphics[width=0.45\textwidth]{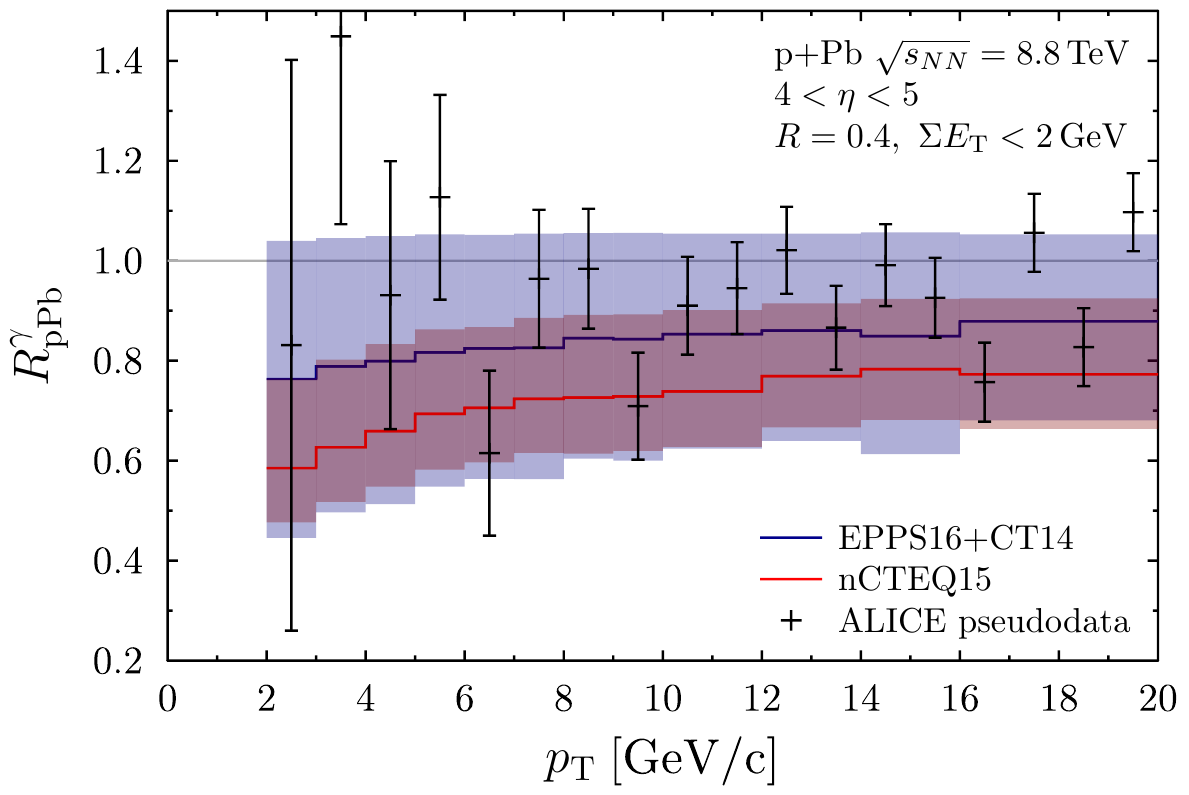}
\vspace{-0.8cm}\caption{The nuclear modification $R_{\rm pPb}$ for isolated photons at forward rapidities as predicted by the EPPS16 and nCTEQ15 nuclear PDFs. Projected pseudodata for the ALICE FOCAL detector are overlaid. Figure by I.~Helenius.}
\label{fig:FOCAL}
\end{wrapfigure}

\vspace{-0.2cm}
A relevant question is what are the options of p--Pb collisions at the LHC to provide further constraints. The potential of open heavy-flavour (D- and B-meson) measurements to offer new information has been recently brought up \cite{Zenaiev:2015rfa,Gauld:2016kpd}. The mass of the heavy quark provides a ``hard'' scale, and it is therefore possible to use perturbative QCD down to zero transverse momentum $p_{\rm T}=0$ and reach sensitivity to very low values of $x$. However, there are theoretical uncertainties related to e.g. the treatment of heavy-quark fragmentation. Also, the multi-parton scattering may play a pronounced at low $p_{\rm T}$ role in collisions involving heavy nuclei \cite{dEnterria:2016yhy}. Bearing in mind such issues, relying on heavy-flavour production in the determination of nuclear gluon PDF may be in doubt. Related theoretical issues are also met in the case of e.g. vector meson production, and generally, in all hadronic observables at low $p_{\rm T}$.

The isolated photon production at forward direction is another observable that has been proposed as a tool for pinning down the nuclear gluon PDF. The advantage is that such an electromagnetic observable is theoretically cleaner in comparison to fully hadronic final states. Figure~\ref{fig:FOCAL} shows the  EPPS16 and nCTEQ15 predictions corresponding to the rapidity domain $4 < \eta <5$ that could be measured by the ALICE FoCal detector \cite{Peitzmann:2016gkt}. The predictions are contrasted with the estimated data precision reachable by the FoCal equipment. Although these measurements should definitely have an impact, for the larger systematic uncertainties at small $p_{\rm T}$, the best constraints appear to be limited to $p_{\rm T} \gtrsim 10 \, {\rm GeV}$. In addition, the probed $x$ region in photon production tends to be shifted to significantly higher $x$ than what a naive leading-order proxy $x \sim (2p_{\rm T}/\sqrt{s}) e^{-\eta}$ would indicate \cite{Helenius:2014qla}. Thus, although the impact of forward-photon measurements has not yet been estimated in detail, the obtainable constraints may be limited at small-$x$ and small $Q^2$.
\begin{wrapfigure}{r}{0.60\textwidth}
\vspace{-1.2cm}
\center
\includegraphics[width=0.61\textwidth]{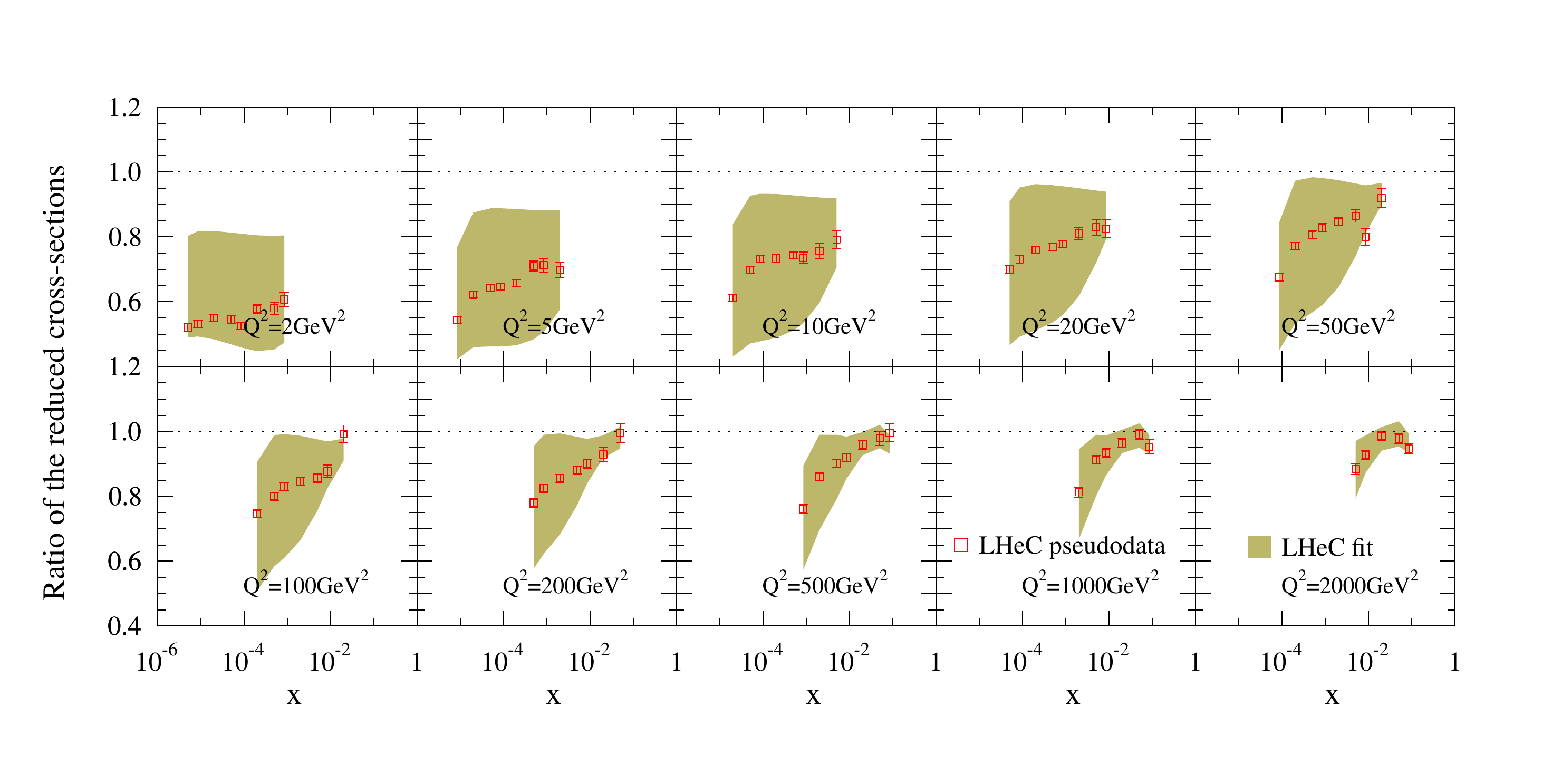} \\
\vspace{-0.7cm}
\includegraphics[width=0.61\textwidth]{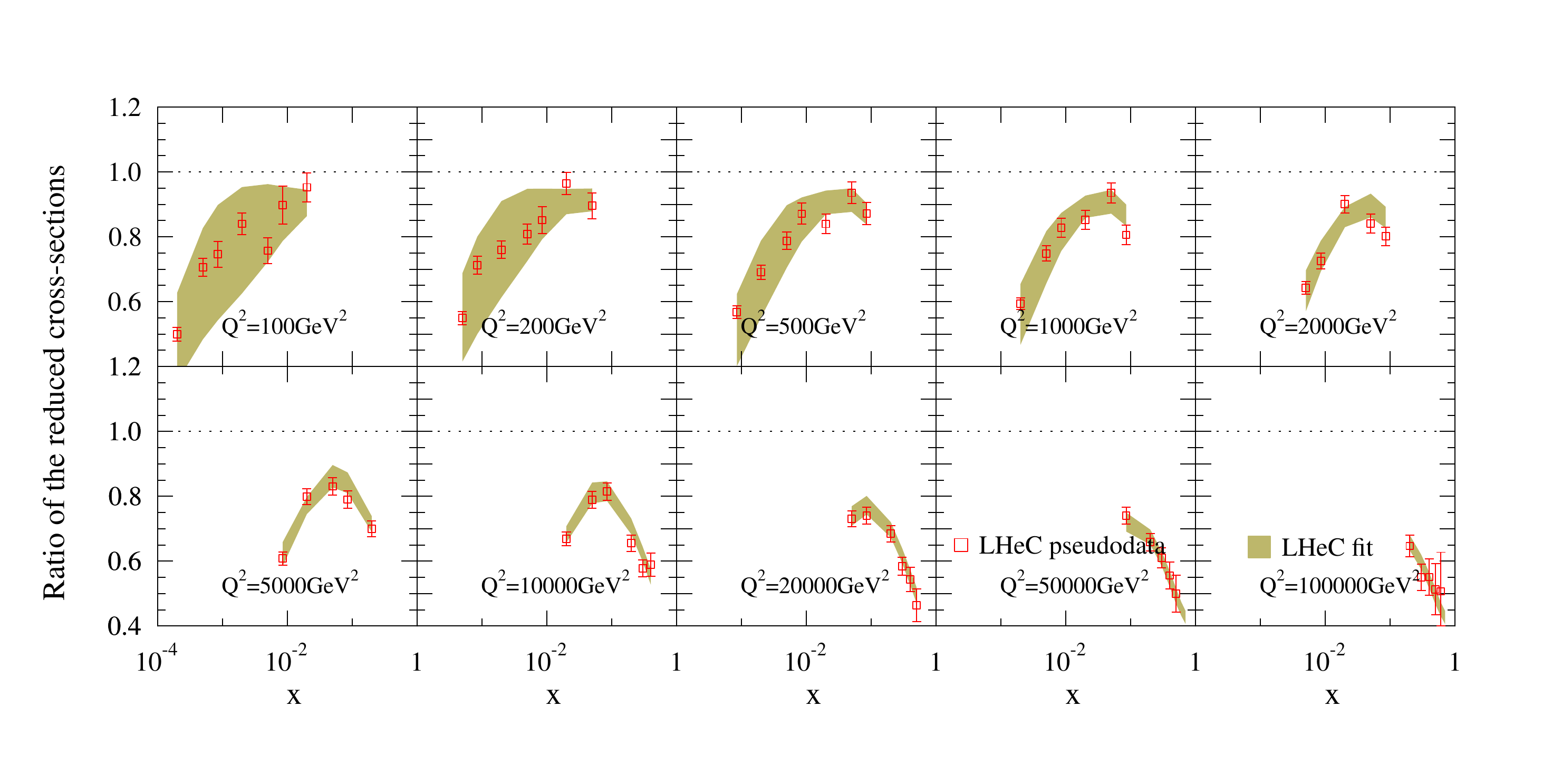} \\
\vspace{-0.6cm}\caption[]{LHeC pseudodata for ratios of reduced NC (upper) and CC (lower) cross sections compared with EPPS16.}
\label{fig:LHeCvsEPPS16}
\end{wrapfigure}

\vspace{-0.8cm}
\section{LHeC and EPPS16}

\vspace{-0.3cm}
To estimate the impact of LHeC measurements, a sample of pseudodata has been generated assuming integrated luminosities $\mathcal{L}_{\rm ep}=10\,{\rm fb}$, $\mathcal{L}_{\rm ePb}=1\,{\rm fb}$ (per nucleon) for collisions of $\sqrt{s_{\rm p}}=7\,{\rm TeV}$ protons and $\sqrt{s_{\rm Pb}}=2.75\,{\rm TeV}$ (per nucleon) Pb ions on $E_{\rm e}=60\,{\rm GeV}$ electrons. The pseudodata are constructed from the reduced cross sections $\sigma^i_{\rm ePb}$, $\sigma^i_{\rm ep}$ (computed with EPS09 nuclear modifications and CTEQ6.6 free-proton PDFs), and the estimated relative point-to-point ($\delta^i_{\rm uncor.}$) and normalization ($\delta^i_{\rm uncor.}$) uncertainties as
$R_{i} = R_i({\rm EPS09}) \times \left[1 + \delta^i_{\rm uncor.} r^i + \delta_{\rm norm.} r^{\rm norm.} \right]$,
where $R_i({\rm EPS09})$ denotes the ratio of cross sections
$R_i({\rm EPS09}) = {\sigma^{i}_{\rm ePb}({\rm CTEQ6.6}\otimes{\rm EPS09})}/{\sigma^{i}_{\rm ep}({\rm CTEQ6.6})}$, 
and $r^i$ and  $r^{\rm norm.}$ are Gaussian random numbers. The effect of these pseudodata have been estimated in the framework of the recent EPPS16 analysis. Figure~\ref{fig:LHeCvsEPPS16} shows how the EPPS16 uncertainties initially compare with the projected LHeC neutral- (NC) and charged-current (CC) pseudodata --- the uncertainties of EPPS16 clearly exceed the data error bars by a large factor.

\begin{figure}[htb!]
\vspace{-0.4cm}
\center
\includegraphics[width=0.83\textwidth]{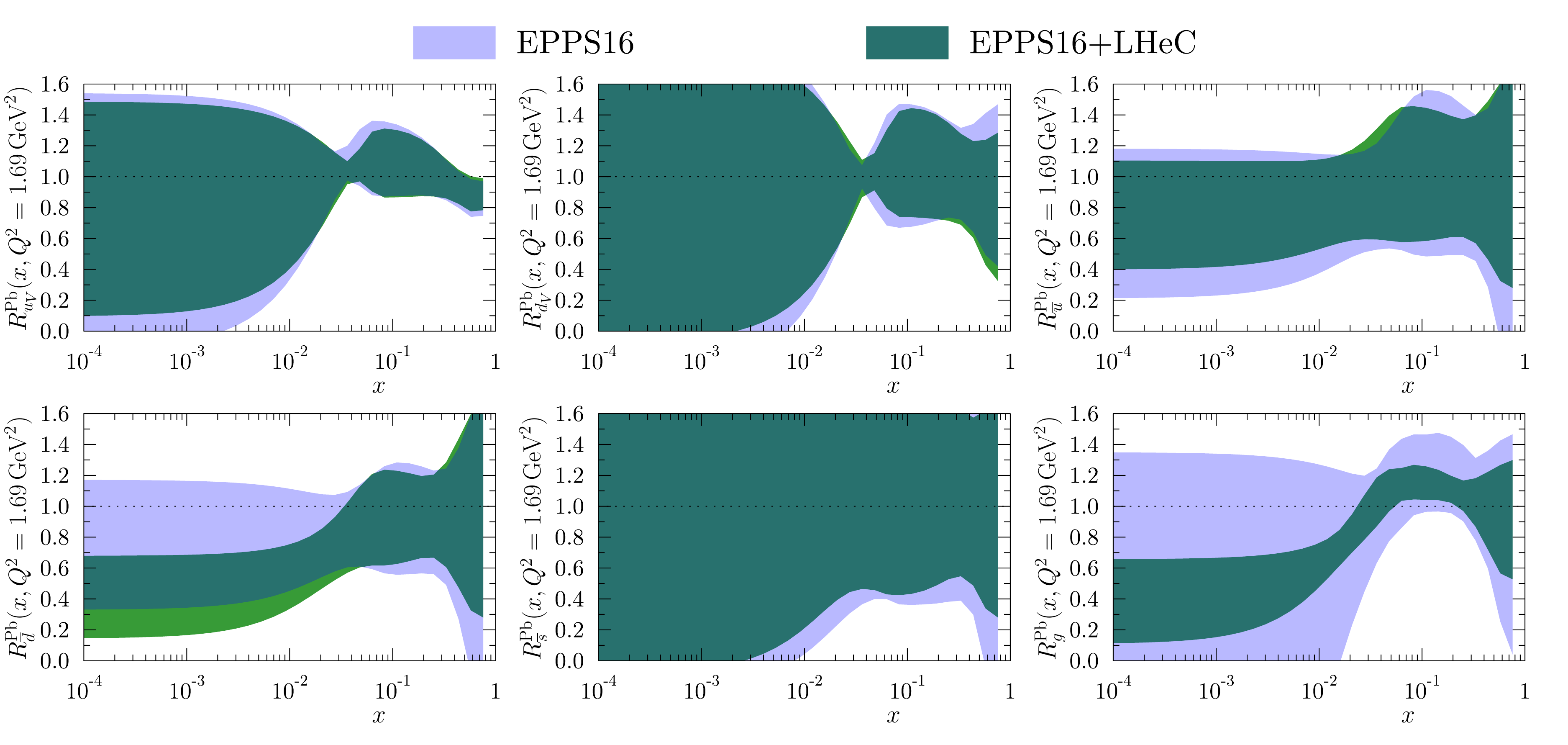}
\vspace{-0.5cm}\caption[]{Nuclear modifications of EPPS16 compared to the EPPS16+LHeC fit at $Q^2=1.69 \, {\rm GeV}^2$.}
\label{fig:neffect}
\end{figure}

The resulting flavour-by-flavour partonic nuclear modifications $R_i^{\rm Pb}$ after including these data in the EPPS16 analysis are shown in Figure~\ref{fig:neffect} at $Q^2=1.69 \, {\rm GeV}^2$, and compared also to the original EPPS16 error bands. For the gluons, the uncertainties go down by a factor of five or so, and especially for $\overline{\rm d}$ distribution the effect is also rather significant. However, it may seem a bit surprising that the uncertainties of the quark sector still remain largish. This can be understood by looking e.g. the valence up-quark distribution $u_V^A$ in the nucleus. For a nucleus $A$ with $Z$ protons, we have
\begin{equation}
u_V^A = ({Z}/{A}) R_{u_{\rm V}} u_{\rm V}^{\rm proton} + \left[({A-Z})/{A}\right] R_{d_{\rm V}} d_{\rm V}^{\rm proton} \,.
\end{equation}
Writing this in terms of the average modification $R_V \equiv ({R_{u_{\rm V}} u_{\rm V}^{\rm proton} + R_{d_{\rm V}} d_{\rm V}^{\rm proton}})/({u_{\rm V}^{\rm proton} + d_{\rm V}^{\rm proton}})$ and the difference $\delta R_{\rm V} \equiv R_{u_{\rm V}} - R_{d_{\rm V}}$, we have
\vspace{-0.2cm}
\begin{equation}
u_V^A = R_{\rm V} \left( \frac{Z}{A} u_{\rm V}^{\rm proton} + \frac{A-Z}{A} d_{\rm V}^{\rm proton}\right) +
        \delta R_{\rm V} \left( \frac{2Z}{A} - 1\right) \frac{u_{\rm V}^{\rm proton}}{1+ u_{\rm V}^{\rm proton}/d_{\rm V}^{\rm proton}} \,.
\end{equation}
The $R_{\rm V}$ term dominates and is indeed very well constrained, see Ref.~\cite{Helenius:2016hcu}. The second term which is sensitive to the flavour separation ($\delta R_{\rm V}$) is always suppressed. However, this is also the case for most of the nPDF applications, and from this viewpoint the remaining uncertainty in flavour separation may not be that critical. Higher luminosities would help especially in the CC case, and lead to a better constrained flavour separation. The addition of charm-production cross section in both NC and CC cases will provide further information on gluon and strange-quark distributions.

\vspace{-0.5cm}
\begin{figure}[!htb]
    \centering
    \begin{minipage}{.30\textwidth}
        \centering
        \includegraphics[width=0.95\textwidth]{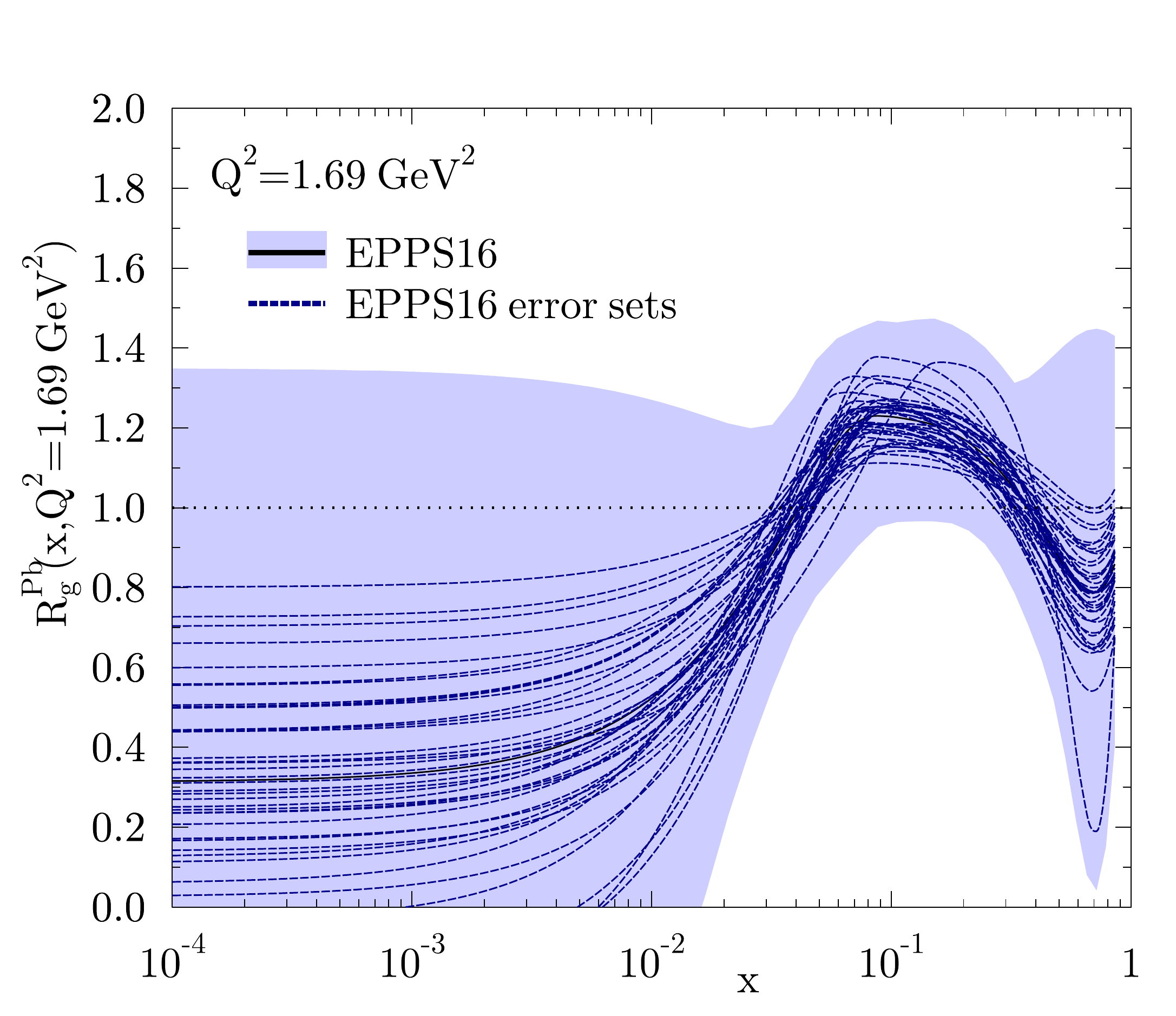} \\
        \includegraphics[width=0.95\textwidth]{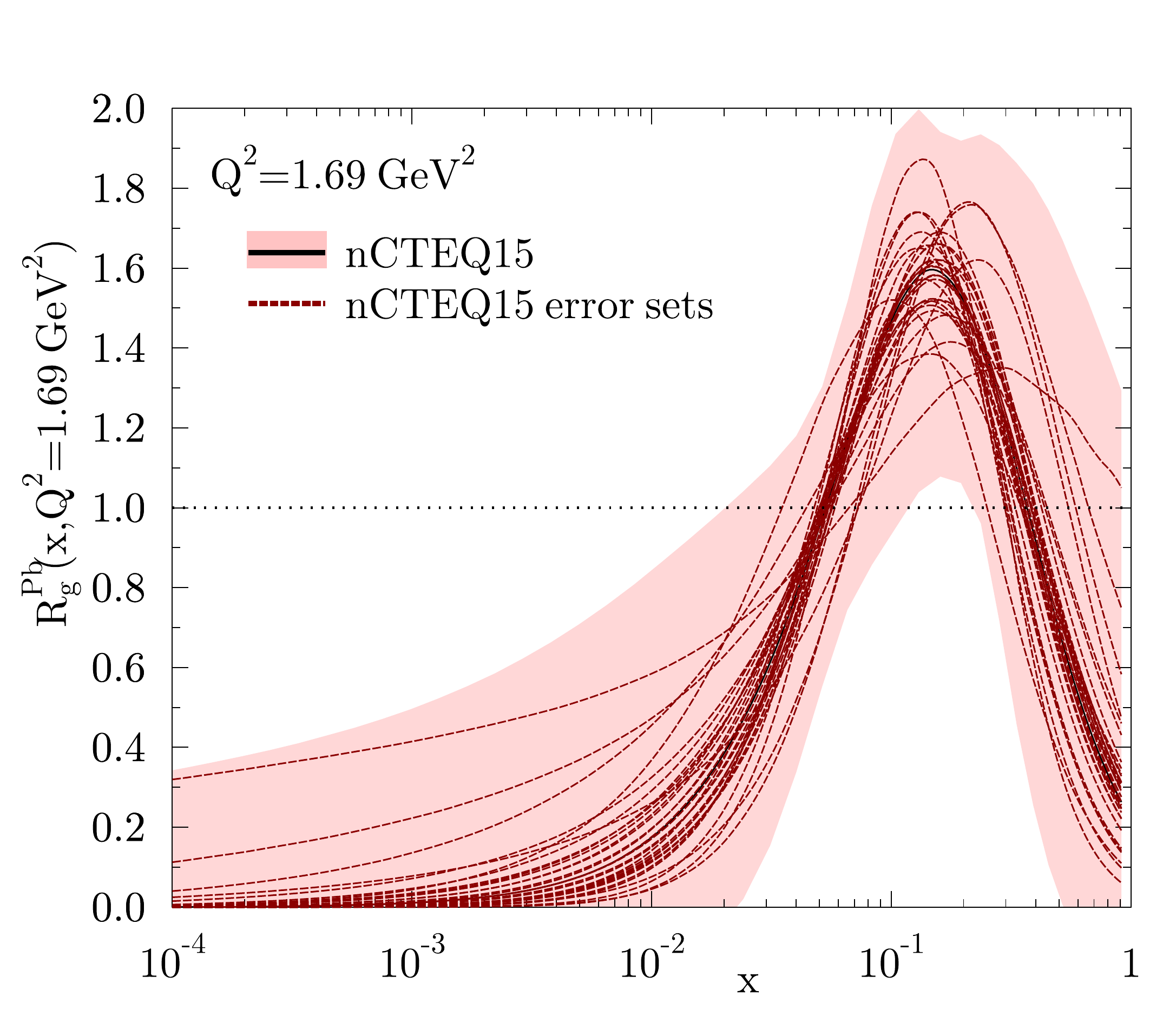}
        \caption{The nuclear modifications for $R_{\rm g}^{\rm Pb}$ at $Q^2=1.69 \, {\rm GeV}^2$ from EPPS16 (upper panel) and nCTEQ15 (lower panel).}
        \label{fig:parbias}
    \end{minipage}%
    \hspace{0.6cm}
    \begin{minipage}{0.60\textwidth}
        \centering
\vspace{0.2cm}
\includegraphics[width=0.43\textwidth]{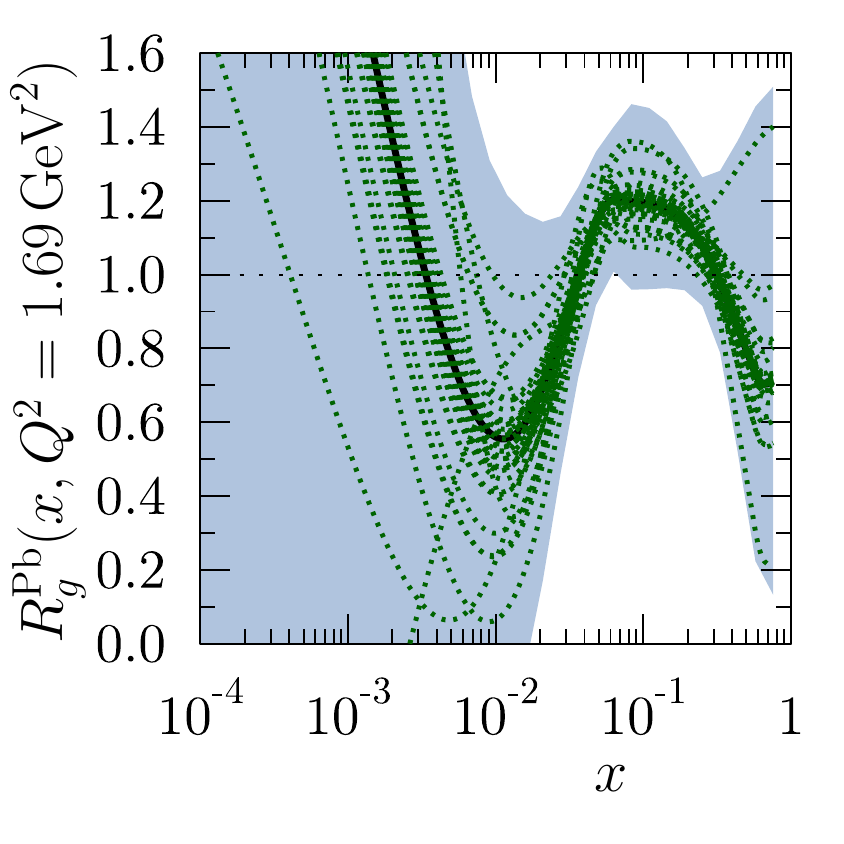}
\includegraphics[width=0.43\textwidth]{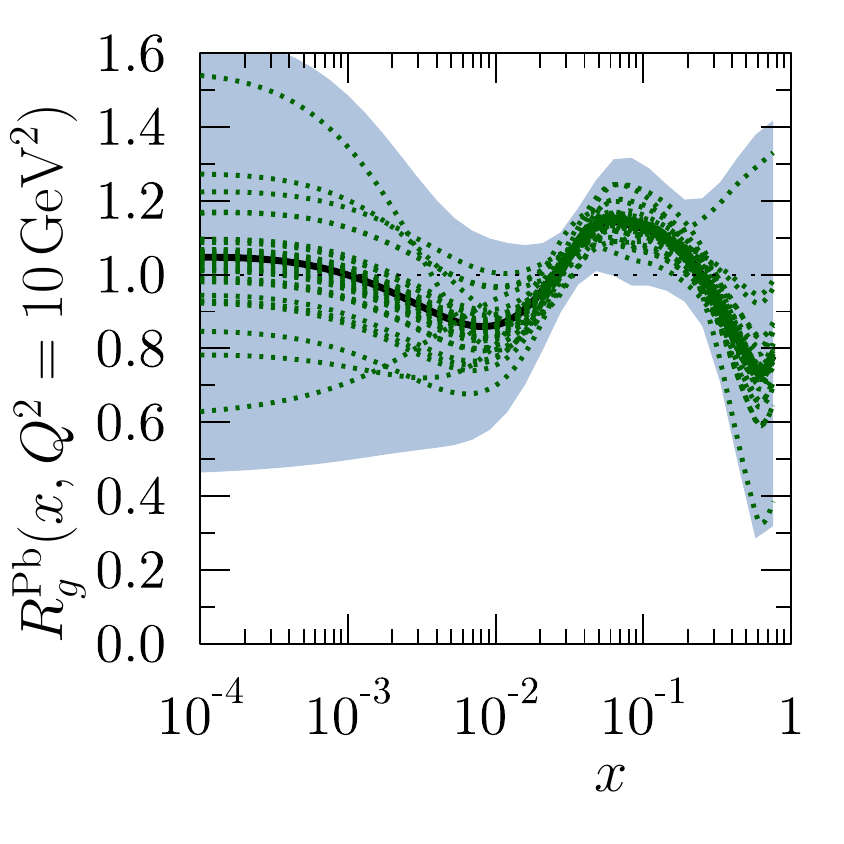} \\
\includegraphics[width=0.43\textwidth]{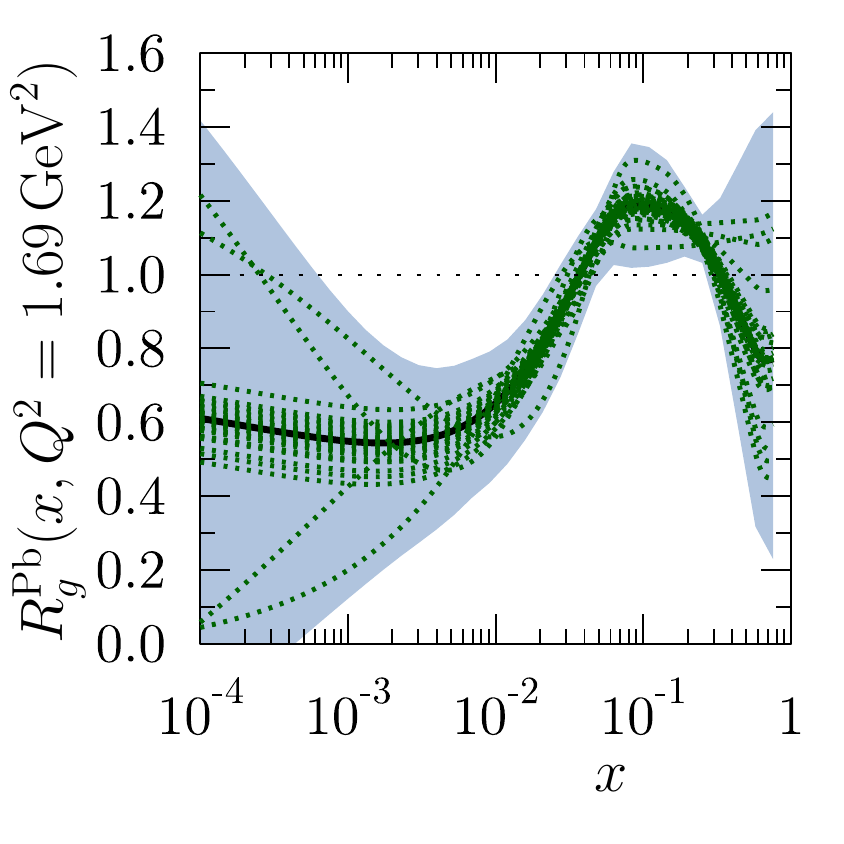}
\includegraphics[width=0.43\textwidth]{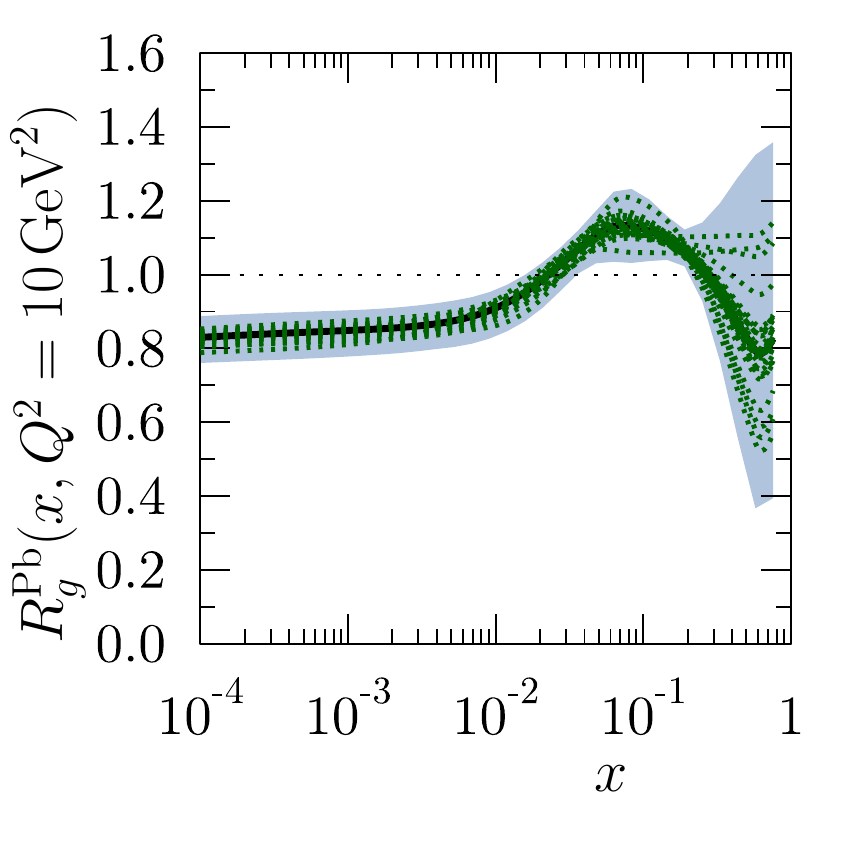}
\vspace{-0.6cm}\caption{The nuclear modification for $R_{\rm g}^{\rm Pb}$ at $Q^2=1.69 \, {\rm GeV}^2$ (left) and $Q^2=10 \, {\rm GeV}^2$ (right) from a fit with increased freedom at small $x$. Upper panels correspond to a fit with no LHeC data, and the lower panels to a fit including the LHeC pseudodata.}
        \label{fig:morefreedom}
    \end{minipage}
\end{figure}

\vspace{-0.7cm}
\section{On the parametrization bias}

\vspace{-0.3cm}
Finally, I would like to comment on the parametrization bias involved in the results shown in Figure~\ref{fig:neffect}. To this end, Figure~\ref{fig:parbias} presents all the error sets $R_{\rm g}^{\rm Pb}$ from the EPPS16 an nCTEQ15 parametrizations. What is notable (alarming?) here is that there is almost no freedom in the functional form at low $x$ --- all the error sets show a monotonic decrease although there are no data constraints. This kind of behaviour is a pure assumption and both, as well as the EPPS16+LHeC fit shown in Figure~\ref{fig:neffect}, underestimate the true uncertainties.

An obvious workaround is to add more flexibility to the small-$x$ fit functions. However, one quickly meets the limitations of the Hessian uncertainty-determination method used in the EPPS16 and nCTEQ15 fits. Particularly difficult is to estimate the uncertainty before the inclusion of the LHeC pseudodata: The Hessian method relies on quadratic expansion of the $\chi^2$ in terms of fit parameters, but when there are no constraints or when they are only very weak, the $\chi^2$ profiles are often very flat with significant higher-order components. In addition, the inter-flavour correlations become very strong. In a situation like the Monte-Carlo methods are superior in comparison to the Hessian method. Despite these difficulties, the Hessian method can give a rough idea if we restrict only to the gluon distributions. Concretely, we have added two extra terms $(x-x_a)^2 \left[a_2 x^\alpha +a_3 x^{2\alpha} \right]$, to the EPPS16 small-$x$ fit function for gluons, where $\alpha=0.25$, and $x_a$ denotes the critical point of antishadowing maximum. The results for $R_g$ at $Q^2=1.69 \, {\rm GeV}^2$ and $Q^2=10 \, {\rm GeV}^2$ are shown in Figure~\ref{fig:morefreedom} before and after the LHeC pseudodata are input. These results give already a better idea how fantastic the impact of LHeC would be on the nuclear gluon densities.

\vspace{-0.4cm}
\section{Summary}

\vspace{-0.3cm}
I have described the present status of the studies concerning the impact of LHeC on nuclear PDFs. The most recent progress has been the inclusion of LHeC pseudodata in EPPS16-based global fits. The inclusive NC cross sections have a significant impact on the gluon and average sea-quark distributions. Perhaps a bit unexpectedly, in the analysis here, the addition of CC cross sections did not lead to a particularly precise flavour separation. In future, this will be improved by using the updated, higher luminosities and considering also charm-tagged observables. A true understanding of the nuclear-PDF uncertainties will require extended small-$x$ fit functions, and I briefly discussed the difficulties this entails in the present EPPS16-like setup.

\vspace{-0.4cm}
\section*{Acknowledgments}

\vspace{-0.3cm}
I have received funding from Academy of Finland, Project 297058; the European Research Council grant HotLHC ERC-2011-StG-279579 ; Ministerio de Ciencia e Innovaci\'on of Spain and FEDER, project FPA2014-58293-C2-1-P; Xunta de Galicia (Conselleria de Educacion) - I have been part of the Strategic Unit AGRUP2015/11.


\vspace{-0.4cm}
\begin{thebibliography}{99}

\vspace{-0.2cm}
\bibitem{AbelleiraFernandez:2012cc}
  J.~L.~Abelleira Fernandez {\it et al.} [LHeC Study Group],
  J.\ Phys.\ G {\bf 39} (2012) 075001
  doi:10.1088/0954-3899/39/7/075001
  [arXiv:1206.2913 [physics.acc-ph]].

\vspace{-0.2cm}
\bibitem{Eskola:2016oht}
  K.~J.~Eskola, P.~Paakkinen, H.~Paukkunen and C.~A.~Salgado,
  Eur.\ Phys.\ J.\ C {\bf 77} (2017) no.3,  163
  doi:10.1140/epjc/s10052-017-4725-9
  [arXiv:1612.05741 [hep-ph]].

\vspace{-0.2cm}
\bibitem{deFlorian:2011fp}
  D.~de Florian, R.~Sassot, P.~Zurita and M.~Stratmann,
  Phys.\ Rev.\ D {\bf 85} (2012) 074028
  doi:10.1103/PhysRevD.85.074028
  [arXiv:1112.6324 [hep-ph]].
  
\vspace{-0.2cm}
\bibitem{Kovarik:2015cma}
  K.~Kovarik {\it et al.},
  Phys.\ Rev.\ D {\bf 93} (2016) no.8,  085037
  doi:10.1103/PhysRevD.93.085037
  [arXiv:1509.00792 [hep-ph]].

 \vspace{-0.2cm}
\bibitem{Hirai:2007sx}
  M.~Hirai, S.~Kumano and T.-H.~Nagai,
  Phys.\ Rev.\ C {\bf 76} (2007) 065207
  doi:10.1103/PhysRevC.76.065207
  [arXiv:0709.3038 [hep-ph]].

\vspace{-0.2cm}
\bibitem{Zenaiev:2015rfa}
  O.~Zenaiev {\it et al.} [PROSA Collaboration],
  Eur.\ Phys.\ J.\ C {\bf 75} (2015) no.8,  396
  doi:10.1140/epjc/s10052-015-3618-z
  [arXiv:1503.04581 [hep-ph]].
  
\vspace{-0.2cm}
\bibitem{Gauld:2016kpd}
  R.~Gauld and J.~Rojo,
  Phys.\ Rev.\ Lett.\  {\bf 118} (2017) no.7,  072001
  doi:10.1103/PhysRevLett.118.072001
  [arXiv:1610.09373 [hep-ph]].

\vspace{-0.2cm}
\bibitem{dEnterria:2016yhy}
  D.~d'Enterria and A.~M.~Snigirev,
  arXiv:1612.08112 [hep-ph].
 
\vspace{-0.2cm}
\bibitem{Peitzmann:2016gkt}
  T.~Peitzmann [ALICE FoCal Collaboration],
  PoS DIS {\bf 2016} (2016) 273
  [arXiv:1607.01673 [hep-ex]].

\vspace{-0.2cm}
\bibitem{Helenius:2016hcu}
  I.~Helenius, H.~Paukkunen and N.~Armesto,
  PoS DIS {\bf 2016} (2016) 276
  [arXiv:1606.09003 [hep-ph]].

\vspace{-0.2cm}
\bibitem{Helenius:2014qla}
  I.~Helenius, K.~J.~Eskola and H.~Paukkunen,
  JHEP {\bf 1409} (2014) 138
  doi:10.1007/JHEP09(2014)138
  [arXiv:1406.1689 [hep-ph]].

  
\end{thebibliography}
\end{document}